\documentclass[conference]{IEEEtran}

\IEEEoverridecommandlockouts
\usepackage{subfigure}
\usepackage{url}
\usepackage{threeparttable}
\usepackage{listings}
\usepackage{multirow}
\usepackage{amsmath}
\usepackage{amssymb}
\usepackage{fancybox, graphicx}
\usepackage{url}
\usepackage{textcomp}
\usepackage{lscape}
\usepackage{booktabs}
\usepackage{colortbl}
\usepackage{paralist}
\usepackage{eurosym}

\usepackage{graphicx}
\usepackage{lipsum}
\usepackage{times}
\usepackage{soul}
\usepackage[utf8]{inputenc}
\usepackage[makeroom]{cancel}

\newtheorem{definition}{Definition}

\AtBeginDocument{%
  \providecommand\BibTeX{{%
    \normalfont B\kern-0.5em{\scshape i\kern-0.25em b}\kern-0.8em\TeX}}}

\usepackage{glossaries}
\newacronym{cps}{CPS}{Cyber-Physical Systems}
\newacronym{cprs}{CPRS}{Cyber-Physical Resilient System}
\newacronym{scada}{SCADA}{Supervisory Control and Data Acquisition}

\usepackage{color}

\usepackage{mathtools}

\begin{document}
\title{Cyber-Resilience Evaluation\\ of Cyber-Physical Systems}

\author{
\IEEEauthorblockN{Mariana Segovia, Jose Rubio-Hernan, Ana R. Cavalli and Joaquin Garcia-Alfaro}
\IEEEauthorblockA{Institut Polytechnique de Paris, T\'el\'ecom SudParis, France}
}

\maketitle
\begin{abstract}
\gls*{cps} use computational resources to control physical process and provide critical services. For this reason, an attack in these systems may have dangerous consequences in the physical world. Hence, resilience is a fundamental property to ensure the safety of the people, the environment and the controlled physical process.
In this paper, we present metrics to quantify the resilience level based on the design, structure, stability, and performance under the attack of a given \gls*{cps}. The metrics provide reference points to evaluate whether the system is better prepared or not to face the adversaries. This way, it is possible to quantify the ability to recover from an adversary using its mathematical model based on switched linear systems and actuators saturation.
Finally, we validate our approach using a numeric simulation on the Tennesse Eastman control challenge problem.\\

\textbf{Keywords:} Cyber-Physical Security, Critical Infrastructures, Resilience, Metrics, Adversary Model, Switched Linear Systems, Actuator Saturation, Networked Control Systems.
\end{abstract}

\IEEEpeerreviewmaketitle

\section{Introduction}
\label{sec:intro}

\noindent Stability refers to the ability of a system to return to the equilibrium point after system disturbances, for example, due to malicious actions that move the system from stable states to unstable ones. Performance aims at working at the desired dynamic response and in a control mode that optimizes the objective function that minimizes costs and maximizes revenues. It must be noted that performance is often calculated after the adversary actions  since it is difficult to know in advance how the system will react to an unknown adversary.

Stability and performance are important factors to accept or reject a system design. The control theory community has provided different criteria to analyze them, such as Lyapunov theory, root-locus, Routh-Hurwitz, Bode or Nyquist methods. These mechanisms are prepared to take into account failure or process errors. However, they are not prepared for malicious actions that may perturb the system.

Historically, malicious actions have not been a system concern since this problem appeared with the introduction of computing resources to control the physical processes. In this context, it is necessary to have mechanisms to provide cyber resilience that goes beyond the traditional failure resilience and it can deal with and correct malicious actions. In addition, it is also necessary to have mechanisms to evaluate at design time the resilience of a \gls*{cps} from a cyber point of view in order to determine its capability to face cyber-physical adversaries.

It is not easy to predict at design time if the system will be stable when facing unknown malicious actions that will be introduced at runtime. However, it is possible to provide reference points to evaluate whether the system is better prepared or not to face the adversaries.

This paper aims at providing a set of metrics that measure at design time the system behavior to determine whether or not it will be acceptable during an attack. We consider both issues: performance and stability. To do so, it may be acceptable to work in a graceful degradation mode while facing an attack, but it must be ensured at least the stability and a minimum performance threshold. We also analyze the internal structure of the system, identifying the critical components that are required to keeps providing its fundamental functions and the capability of the system to restore the crucial components in case of damage due to attacks.

Consequently, the main contributions of this paper can be summarized as follows: (1) we provide a mechanism to evaluate at design time the resilience of a \gls*{cps} in the presence of cyber-physical adversaries considering both the performance and stability of the system, and the design and its structure; (3) we sum up guidelines to improve the resilience by design of a \gls*{cps}; and (4) we provide experimental work to validate the approach.

\medskip
\noindent \textbf{Paper Organization ---} Section~\ref{sec:background}
provides the related work. Section~\ref{sec:assumptions} provides
preliminaries and assumptions about the system and the adversary. Section~\ref{sec:metrics} presents our resilience metrics to evaluate it. Section~\ref{sec:experimental} reports
the experimental work and Section~\ref{sec:conclusion} concludes the
paper.


\section{Related Work}
\label{sec:background}

Resilience is an important system's property that has been analyzed in different disciplines and contexts in the last years. In this section, we analyze different evaluation mechanisms from a cyber and cyber-physical point of view.

In order to achieve resilient systems, it is important to develop appropriate metrics to assess it and demonstrate the utility of the proposed approaches.
Different research works have identified attributes to measure the resilience of a system. In the sequel, we present the main findings.

In \cite{fang_resilience-based_2016}, authors propose a metric to evaluate the criticality of a component in a network system from the perspective of their contribution to resilience. Specifically, the two proposed metrics quantify the priority with which a failed component should be repaired and the potential loss in the optimal system resilience due to a time delay in the recovery of a failed component. In this paper, we analyze the resilience of the system as a whole. The objective is to quantify whether a proposed approach improves resilience or not.

In \cite{francis_metric_2014}, the authors proposed a resilience analysis framework and a metric for measuring it. The framework is focused on the achievement of three resilience capacities: adaptability, absorbability, and recoverability. These properties are the basis for the resilience metric.
This  approach presents a general metric designed to apply to a wide variety of systems, such as physical, economic, social, ecological, among other types of systems. Due to its generality, this mechanism is not the most suitable for evaluating the reaction of a \gls*{cps} when facing an attack, since it is not capable of considering all the specific characteristics of this kind of system.

A metric is provided in \cite{linkov_resilience_2013} that describes resilience in four dimensions on a policy level. However, it does not capture the runtime performance of a system or the temporal component of the resilience, which its an important factor to consider.

In \cite{rieger_resilient_2014} is presented a metric framework that integrates the cognitive, cyber, and physical aspects considering time and data integrity characteristics. The resilience is considered with respect to control stability and the author uses control response and stability as a performance measure. Similarly, \cite{eshghi_power_2015} models a system as a hierarchical set of controllers that are then behavior bottom-up to retrieve system resilience. The proposed approaches
use traditional performance metrics to provide a visualization methodology for operators and indications of issues that show the impact of the disturbances. These metrics are related to state awareness of the real-time operation, but they do not allow to evaluate in advance the reaction of the system. In addition, they consider physical threats and cyber threats in a separate manner. Hence, it is not clear if the approach will be able to handle cyber-physical adversaries capable of making the system to lose the state monitoring of the system.

In addition, the authors in \cite{clark_cyber-physical_2019} propose a resilience metric for \gls*{cps} modeled as linear systems with and without actuator saturation. It considers both the physical and cyber aspects of the systems. They quantify the ability of the system to recover from an attack under the assumption that the attack is discovered within a fixed time interval and evaluating its domains of attraction. The proposed physical evaluation is based on the stability evolution of the system. However, it is a mathematical abstract definition that may be hard to apply to practical evaluation.

In this paper, we provide a metric specific for \gls*{cps} systems that takes into account the temporal dimension of resilience. It is based on the stability and performance of the physical process in order to guarantee that the required safety properties are met. The objective is to provide a mechanism to assess the resilience of the system at the design time.

\section{Preliminaries}
\label{sec:assumptions}

We provide in this section our assumptions about the system and the adversary models as well as some initial preliminary concepts.

\subsection{System Model}
\label{subsec:sys_model}
A cyber-physical system can be mathematically modeled as follows:
\begin{equation}
  \label{eq:ch3_state}
  x_{k+1}=Ax_{k}+Bsat(u_{k}) + w_{k}
\end{equation}
where $x_{k}\in \mathbb{R}^n$ is the vector of the state variables at the $k$-th time step, $u_{k}\in \mathbb{R}^p$ is the control signal, and $w_{k}\in\mathbb{R}^n$ is the {\it process noise} that is assumed to be a zero-mean Gaussian white noise with covariance $Q$, {\it i.e.} $w_k \sim N(0,Q)$. Moreover, $A\in \mathbb{R}^{n\times n}$ and $B\in \mathbb{R}^{n\times p}$ are respectively the {\it state} matrix and the {\it input} matrix.

Actuator saturation is an inherent non-linearity feature in dynamic systems caused by constraints that reflect bounds or limits in actuators. The saturation function $sat: \mathbb{R} \rightarrow \mathbb{R}$ is defined as follows: $ sat(u_i) = sign(u_i) min \{ |u_i|, S_{max} \}$ where $u_i$ is one entry of the command input $u$ that is calculated as $u = Kx$ with $K$ the feedback gain matrix and $S_{max}$ is the maximum saturation level. For a vector $u \in \mathbb{R} ^m$ we define $sat(u)$ as $sat(u) = [sat(u_1) sat(u_2) ... sat(u_m)]$.

The saturation limits the maximum command that may be executed at every time step. Hence, it is important to consider the actuator saturation for resilience because it limits the impact of the adversary on the system \cite{kafash_constraining_2017}, but it also limits the response of the system to recover due to its implications on the stability and reachability of control.
Actuators can not inject arbitrarily large amounts of energy into the system since there are always physical limitations and the saturation arises from these limits.

A static relation maps the state $x_k$ to the system output $y_k \in \mathbb{R}^m$:
\begin{equation}
  \label{eq:ch3_output}
  y_{k}=Cx_{k}+v_{k}
\end{equation}
where $C\in \mathbb{R}^{m\times n}$ is the output matrix. The value of the output vector $y_{k}$ represents the measurement produced by the sensors that are affected by a noise $v_{k}$ assumed as a zero-mean Gaussian white noise and covariance $R$, {\it i.e.} $v_k \sim N(0,R)$.

We assume a system that is stable and showed optimal control under normal conditions (i.e. in the absence of malicious actions).

\subsection{Adversary Model}
\label{subsec:adv_model}

The objective of the adversary is to cause a malfunction in the system by performing actions that affect the control system. The adversary is situated in a remote location but gained access to the internal network exploiting some cyber vulnerabilities and uses the network traffic to perform the attack as an insider.

A cyber-physical adversary can be modeled mathematically as follows

\begin{equation}
  \label{eq:att_1}
  x'_{k+1}=Ax_{k}+B' sat(u'_{k}) + w_{k}
\end{equation}
\vspace{-.5cm}
\begin{equation}
  \label{eq:att_2}
  y'_{k}=C'x_{k}+ v_{k}
\end{equation}

where $B'sat(u'_{k})$ represents an attack to the control input. The matrix $B'$ is estimated by the attacker for the system model matrix B and $u'_{k}$ is a malicious command.  $C'$ represents an attacker that is able to create a malicious sensor outputs $y'_{k}$. These malicious actions may be done by compromising sensors, actuators, controllers or network links.

The adversary that corrupts the system measurements and the command outputs simultaneously is the most powerful adversary, which is the parametric cyber-physical adversary described in \cite{rubio2017EurasipWatermak}. The adversary has the ability to estimate the system parameters, for example, with techniques such as machine learning, ARX (\textit{Autoregressive with exogenous input}) or ARMAX (\textit{Autoregressive-moving average with exogenous input}) models.

 In addition, this adversary can inject specific malicious measurements designed to deceive the control system as in Equation \ref{eq:att_2} using the matrix $C'$ which is also an estimated of the real system parameter. This means that the attacker will try to send a sensor output according to the system state $x_k$ that the controller is expecting. This attack is designed intentionally to mislead the system or destabilize it without being detected. In opposite to faults that have a random nature and are much easier to be detected and mitigated. The closer the matrices $B'$ and $C'$ are to the real matrices $B$ and $C$, the more dangerous is the adversary.

\section{Resilience metrics}
\label{sec:metrics}

This section presents the metrics to evaluate the system resilience at design time.

\subsection{Performance and Stability Analysis}
\label{subsec:stab_perf}

This analysis determines whether the system will remain stable and meet the minimum performance threshold under attack. It allows quantifying the maximum time that the system can resist in the absorb phase under attack, i.e., the maximum time that it has to react and stop the malicious actions. It also allows determining the states that the system may reach during the malicious action and estimate the maximum performance damage that the adversary may cause.

Traditionally, the performance is used to measure the deviation between the process dynamics and the models to control it. In addition, it can be used to evaluate the resilience of a system by analyzing the capacity to absorb and recover from malicious action. In this section, we evaluate the underlying physical model to dimension the maximum performance loss during the worst attack scenario.

\medskip

\noindent \textbf{Thresholds and Setpoints:} The performance must be defined according to the defined process operating objectives. For example, some possible objectives are safety conditions, product quality, environment protection, equipment protection, quality control, profit, among others. These established objectives will define process constraints that can be expressed as restrictions over the state of the system and they can be controlled through the monitored process variables. These restrictions define the performance thresholds ($TS$) that must be satisfied even when the system is working under attack. Hence, the first step is to establish the minimum performance threshold that is required, and the setpoint ($SP$) for the normal system behavior.

In addition, the performance should be evaluated over a period of time which we will divide in the absorb and recover phase. The absorb phase starts with the attack in time $k_0$ and finishes in time $k_a$ when the system reaches its minimum performance. The recovery phase starts in $k_a$ and finishes in $k_r$ when the system recovers its normal performance in the setpoint $SP$. This allows estimating the maximum derivation during the attack to evaluate if the performance threshold will be ensured. A small state variation during the attack is desirable so that the process variable remains close to its equilibrium state.

The resilience is based on the absorbing and recovering potential. The absorbing property of a system is the degree to which challenges can be handled even with performance degradation. The recovery potential describes a system's ability to restore normal operation in the face of challenges. To estimate the performance, we will evaluate the system evolution during the absorb and recover phases.

\medskip

\noindent \textbf{Absorb Phase Time:} The absorb time ($KA$) corresponds to the time required for the resilience approach to start working. In particular, the system is defined as resilient if for any adversarial input in the absorb phase the resulting state is within the threshold range.

\medskip

 \noindent \textbf{Recover Phase Time:} The recovery time ($KR$) which corresponds to the period $k_r - k_a$ depends on how fast the system can be stabilized. It can be estimated with the settling time of the control system. The settling time is defined as the time taken for the process response to settle within near a constant value, usually in some band within $2\%$ around the equilibrium state \cite{marlin_process_2000}.

\medskip

\noindent \textbf{Maximum Deviation:} The maximum deviation ($MD$) of the controlled variable from the $SP$ is an important measure of the process degradation. We assume that in normal behavior the system is in the $SP$. Hence, the maximum deviation $MD$ corresponds to the difference between the $SP$ and the possibles deviations $\Psi$ during the attack.

$$
MD = max\{|\Psi - SP|\}
$$

Given the time $KA$, calculated as in the previous section, which is the maximum time the system can be in the absorb phase, it is possible to calculate the states that can be reached in the worst case scenario where the attacker takes the system to its saturation level.

Hence, the maximum reachable state in time $KA$ is calculated by substituting recursively the state $x_k$ in the period $k_0$ and $k_a$ as follows

$$
\chi = x_{k_a} = A x_{(k_a-1)}  \pm B S_{max}
$$
\begin{equation}
  \label{eq:chi}
    \chi = A^{KA} SP \pm \sum\limits_{i=1}^{KA} A^{(KA - i)} B S_{max}
\end{equation}
where $\chi$ indicates the maximum and minimum reachable states using the saturation level $S_{max}$. The set $\Psi$ can be determined as $\Psi = C\chi$ using Equation \ref{eq:ch3_output}.

\medskip

 \noindent \textbf{Resilience Loss:}  The resilience loss ($RL$) is the sum of the differences between $SP$ and the actual performance of the monitored variables during the absorb and recovery phase, i.e., $RL$ is the sum of areas above and below the setpoint.

 \vspace{-0.2cm}
 \begin{equation}
  \label{eq:rl}
    RL = (\sum_{j=k_0}^{k_r} |y_j - SP|)
\end{equation}

 \noindent \textbf{Performance and Stability Resilience:}  The Performance and Stability Resilience $PR$ can be estimated as the area defined within the thresholds $TS$ during the absorb and recovery phase less the resilience loss $RL$.
 \begin{equation}
  \label{eq:res}
    PR = (NR - RL) / NR
\end{equation}
where $NR = (TS_{sup} - TS_{inf}) \times (KA + KR)$.

\medskip

\noindent \textbf{General Performance and Stability Evaluation:} The \textit{performance and stability resilience analysis} quantifies the impact of the attack in one of the monitored variables. For this reason, it is desired to have an overall metric that contemplates the global state of the system.

Not all the components contribute equally to develop the system crucial functions. Hence, not all resources are equally likely to be used by an attacker. The resource's contribution to a system's attack surface depends on the resource's damage potential, i.e., the level of harm the attacker can cause to the system in using this resource in an attack. The higher the damage potential, the higher the contribution to the attack surface.

In addition, the resilient must be evaluated considering the process operation objectives. As we mentioned previously, these objectives establishes the process constraints that create state and monitored variables restrictions.
For this reason, we need to evaluate the resources that are part of the system's attack surface to determine whether they are critical from the objectives point of view. Then, it is possible to define $c^{AT}_j$ as the contribution of the monitored variable $j$ to the attack surface according to the defined objectives.

We calculate the global performance and stability resilience ($GR$) index by pondering the performance and stability resilience $PR$ evaluation of each measured variable $j$ according to their contribution to the attack surface as follows:
\begin{equation}
  \label{eq:rl}
    GR = \sum_{j=1}^{m} c^{AT}_j \times min(PR_j)
\end{equation}

\subsection{Design and Structure Analysis}
\label{subsec:design_struct}

To create a resilient \gls*{cps}, it is required to build a set of stable models to activate when facing an attack. In this section, we will review the techniques proposed in the literature to achieve resilient designs and how to evaluate its structure according to the adversaries the system can recover from.

The cyber-physical adversaries compromise the process by affecting the ability to maintain situational awareness of the process (i.e. affecting the observability) or by reducing the ability to bring the process to the desired state (i.e. affecting the controllability), or a combination of both.


\begin{definition}[Controllability \cite{Kalman60contributionsto}\cite{KALMAN1960491}]
\label{def:controllability}
A system is controllable if every state vector $x_k$ can be transformed into the desired state in finite time by the application of control inputs $u_k$. The controllability depends only on matrices A and B  since a necessary and sufficient condition for a system to be controllable is that the controllability matrix $\mathfrak{C} (A, B)$ has $n$ linearly independent columns.

\begin{equation}
  \label{eq:mtxC}
    rank \ \mathfrak{C} (A,B) = rank [B|AB|...|A^{n-1}B] = n
\end{equation}
\end{definition}


\begin{definition}[Observability \cite{Kalman60contributionsto}\cite{KALMAN1960491}]
\label{def:observability}
The system is observable in $n$ time-steps when the initial state $x_0$ can be recovered from a sequence of observations $y_0,..., y_{n-1}$ and inputs $u_0,..., u_{n-1}$. The observability depends only on matrices A and C  since a necessary and sufficient condition for a system to be observable is that the observability matrix $\mathfrak{O} (A, C)$ has $n$ linearly independent rows.

\begin{equation}
  \label{eq:mtxO}
    rank \ \mathfrak{O} (A, C) = rank \begin{bmatrix}C \\ CA \\ ... \\ CA^{n-1}\end{bmatrix} = n
\end{equation}
\end{definition}

As mentioned previously, there are four components in the attack surface that may be attacked: sensors, actuators, controllers, and network traffic. For this reason, the generated models should address the vulnerabilities exploited in one or more of these components. The resilience design of a \gls*{cps}  can be characterized by the actuator resilience $R_A$, the sensor resilience  $R_S$, the control resilience  $R_C$, and the communication resilience  $R_N$. \\

\begin{definition}[Actuator Resilience  $R_A$]
\label{def:k-act_res}
A \gls*{cps} is t-actuator resilient if $ rank \ \mathfrak{C} (A, B^\Gamma) = n$, i.e. the system is controllable for all possible subset $\Gamma$, where $\Gamma$ is the set of all  possible combinations of actuators removing $t$ critical compromised actuators.
\end{definition}

\begin{definition} [Sensor Resilience  $R_S$]
\label{def:k-sen_res}
A \gls*{cps} is t-sensor resilient if $rank \ \mathfrak{O}(A, C^\Delta) = n$, i.e. the system is observable for all the subsets in $\Delta$, where $\Delta$ is the set of all possible combinations of sensor removing $t$ critical compromised sensors.
\end{definition}

This means that the system will be resilient if the controller can take action despite the compromised parts of the system. The definition of the matrices $B^\Gamma$ and $C^\Delta$ depends on the particular resilience strategy applied to improve the actuator or the sensor resilience.\\

\begin{definition} [Control Resilience  $R_C$]
\label{def:k-ctl_res}
A \gls*{cps} is t-control resilient if $ rank \ \mathfrak{C}(A, B^\Lambda) = n$ and $rank \ \mathfrak{O}(A, C^\Lambda) = n$, i.e, the system is controllable and observable for all possible subset in $\Lambda$ which is obtained by removing $t$ possible compromised critical controllers.
\end{definition}

This definition means that if t-critical-controllers are compromised, the system can keep working and recover the state to an equilibrium point without these controllers working.\\

\begin{definition}[Communication Resilience  $R_N$]
\label{def:k-ntw_res}
A \gls*{cps} is t-communication resilient if $ rank \ \mathfrak{C} (A, B^\Gamma) = n$ and $rank \ \mathfrak{O}(A, C^\Gamma) = n$, i.e, the system is controllable and observable for all possible subset in $\Gamma$ removing $t$ compromised network links in which the adversary has the ability to recover the system model from collected data. \\
\end{definition}

Next, we review different strategies that allow to improve the resilience of a \gls*{cps} in each of its dimensions. We start from a minimum \gls*{cps} with no resilience and progressively increase it with different techniques.

The minimal possible configuration is a \gls*{cps} with the minimum amount of actuators and sensors to work, an automated controller capable of correcting errors in the process, and a non-redundant network that provides connectivity. This basic system provides observability and controllability in order to ensure fault correction. However, it is not resilient to attacks.

To achieve a resilient \gls*{cps} is required to improve the system design including, for example, techniques as the following ones.

The \textit{actuators resilience} $R_A$ can be improved by adding diversified actuators to perform the control actions over the system. Another proposal to improve the actuator resilience is presented in \cite{kanellopoulos_moving_2019} which provides a resilient approach based on moving target defense techniques that use this principle to protect \gls*{cps}s from actuator and sensor attacks. In addition, in \cite{fawzi_secure_2014}, authors define a decoder that can also correct attacks in actuators or sensors that have been corrupted.
The strategies to improve actuator resilience require to add extra hardware devices that help to compensate for the incorrect function of the affected ones.

The \textit{sensor resilience} $R_S$ can be improved using different techniques. Firstly, in a similar way as the previous case, it is possible to add a diversified sensor. In addition, sensor resilience can be improved using software approaches that do not require to add extra hardware devices. For example, the techniques proposed in \cite{han_towards_2016}, \cite{corradini_robust_2017},  \cite{combita_mitigating_2019} and  \cite{pajic_attack-resilient_2017} provide resilient state estimation and reconstruction in the presence of integrity attacks.

Another software approach to improve the sensor resilience is to use an auxiliary system with Luenberger observers \cite{schellenberger_detection_2017}.

The \textit{controller resilience} $R_C$ can be improved by adding local capabilities in the devices, for example, a smart actuator with an embedded local controller that can take control decisions outside the domain of the adversary.
Another option is to implement distributed controllers that implement voting techniques to reach consensus to avoid malicious nodes. This problem has been studied extensively in distributed computing \cite{Lamport_byzantine} \cite{Fekete_consensus}.
Also, techniques such as secret sharing \cite{Shamir_secret} \cite{Brickell_secret} \cite{beimel_secret} and distributed trust \cite{Abdul_distribTrust} \cite{Josang_distribTrust} may be used to implement, for example, mechanisms that divide the control into shares, such that the system needs to reach a given threshold prior granting control. Below the threshold, the information gets concealed from the eyes of the adversary.

The \textit{communication resilience} $R_N$ can be addressed as a problem of transmitting information in the presence of misbehaving nodes has been widely studied in communication networks \cite{Hromkovic_networks} \cite{leblanc_resilient}.
To improve the network resilience one possibility is to add redundant physical or virtual independent networks.
Other mechanisms such as \cite{5605238} showed that linear iterative strategies are able to achieve the minimum bound required to disseminate information reliably, so malicious nodes will be unable to prevent from calculating any function (under a broadcast model of communication).
Finally, in \cite{cyberICPS_reflective}, it is proposed a dynamic mechanism that  dynamically create auxiliary controllers that help the switches to sanitize the traffic modified by the adversary in the network exchange.

\section{Experimental Work -- Numeric Simulation}
\label{sec:experimental}

In this section, we analyze whether a system is resilient using the proposed metrics. To validate the approach, we estimate the defined metrics using a Matlab numeric simulation with a simplified version of the Tennesse Eastman (TE) control challenge problem~\cite{ricker1993model}.
The physical process consists of an isothermal reactor with a separation system. In it occurs an irreversible reaction where the reactants A and C generate the product D. The reaction rate depends only on the partial pressures of A and C.

\medskip

\noindent \textbf{Manipulated Variables:} The control objective is to maintain the product flow rate at a specified value by manipulating the flows of two feeds steams, one purge stream, and the liquid holdup volume.

The two controlled feeds to the reactor chamber are Feed 1 and Feed 2. Feed 1 ($u_1$) consists of the reactants A and C, and traces of an inert gas B. Feed 2 ($u_2$) consists of pure A, which is used to compensate for disturbances in the partial pressures of A and C in Feed 1.

The purge rate ($u_3$) depends on the pressure in the vessel and the position of the purge control valve. The vapor phase can be assumed to consist only on A, B, and C, and the liquid, pure D.

The product flow rate ($u_4$) is adjusted using a proportional feedback controller which responds to variations in the liquid inventory. The regulatory control problem is to maintain a specified product rate by manipulating flows of streams 1, 2, and 3.

\medskip

\noindent \textbf{Controlled variables:} The monitored variables are the production rate (F4), the pressure (P), the liquid inventory (VL) and the amount of reactant A in the purge flow (yA3).

\medskip

\noindent \textbf{Physical Model:} The system is described by the following matrix of transfer functions.
\begin{equation}
\label{transferFunc}
    y=\begin{bmatrix}
      F4    \\
      P     \\
      yA3   \\
      VL
    \end{bmatrix}=Gu=
    \begin{bmatrix}
      g_{11}    & 0         & 0       & g_{14}      \\
      g_{21}    & 0         &g_{23}   & 0           \\
      0         & g_{32}    & 0       & 0           \\
      0         & 0         & 0       & g_{44}
    \end{bmatrix}\begin{bmatrix}
      u_1    \\
      u_2    \\
      u_3    \\
      u_4
    \end{bmatrix}
\end{equation}

The individual transfer functions are given below (the unit of $s$ is seconds).

$$
\begin{matrix}
    g_{11} =
    \dfrac
        {0.02833}
        {45s+1}
    &
    g_{21} =
    \dfrac
        {45(340s+1)}
        {9000s^2+615s+1}
    \\\\
    g_{23} =
    \dfrac
        {-900s-11.25}
        {9000s^2+615s+1}
    &
    g_{32} =
    \dfrac
        {1.5}
        {600s+1}e^{-6s}
    \\\\
    g_{14} =
    \dfrac
        {-3.4s}
        {360s^2+66s+1}
    &
    g_{44} =
    \dfrac
        {1}
        {60s+1}
\end{matrix}
$$

\subsection{Resilience Evaluation}
\label{subsec:evaluation}

The first step to evaluate the resilience of a system is to determine the system threshold, the setpoints, and the saturation limits for its variables.
These parameters are determined by the restrictions from the physical aspects of the plant. We used the data provided in the TE problem~\cite{ricker1993model}. Tables \ref{tab:man_var} and \ref{tab:mes_var} summarize the manipulated and measured variables.

\begin{table}[!b]
\begin{center}
\begin{tabular}{| c | l | l | c |}
\hline
Variable & Input for setpoint & Description & Saturation \\ \hline
u1 & 60.95327313484253 & Feed 1 valve position & 0–100\%\\ \hline
u2 & 25.02232231706676 & Feed 2 valve position & 0–100\%\\ \hline
u3 & 39.25777017606444 & Purge valve position & 0–100\%\\ \hline
u4 & 44.17670682730923 & Liquid inventory setpoint & 0–100\%\\ \hline
\end{tabular}
\medskip
\caption{Manipulated variables \cite{ricker1993model}.\label{tab:man_var}}

\begin{tabular}{| c | l | l | c | c |}
\hline
Variable & setpoint & Description & Units & Threshold\\ \hline
F4 & 100.00 & Product flow  & kmol/hr & - \\ \hline
P & 2700.00 & Pressure &kPa & 2k - 3k\\ \hline
VL & 44.18 & Liquid inventory & \% & 0 – 100\\ \hline
yA3 & 47.00 & Amount of A in purge & mol \% & 0 – 100\\ \hline
\end{tabular}
\medskip
\caption{Controlled variables \cite{ricker1993model}.\label{tab:mes_var}}

\end{center}
\end{table}

The physical process objective is to maximize the production rate while keeping a safe state.

\medskip

\noindent \textbf{Thresholds:} The system thresholds are expressed in Table \ref{tab:mes_var}. In particular, the operating pressure must be kept below 3k Pa due to safety restrictions. Otherwise, the system should be shutdown.

\medskip

\noindent \textbf{Saturation limits:} The limits for each actuator are in Table \ref{tab:man_var}. The flow rates saturate at some point and each valve can variate in a range of 0 to 100 \% open to variate the flow rate.

\medskip

\noindent \textbf{Setpoints:} The setpoints are in Table \ref{tab:mes_var}. In addition, in the column Input for SP in Table \ref{tab:man_var} are expressed the input associated with those setpoints.

\begin{figure*}[htb]
\centering
  \subfigure[\label{fig:max_no_res}]{
  \includegraphics[width=0.45\linewidth]{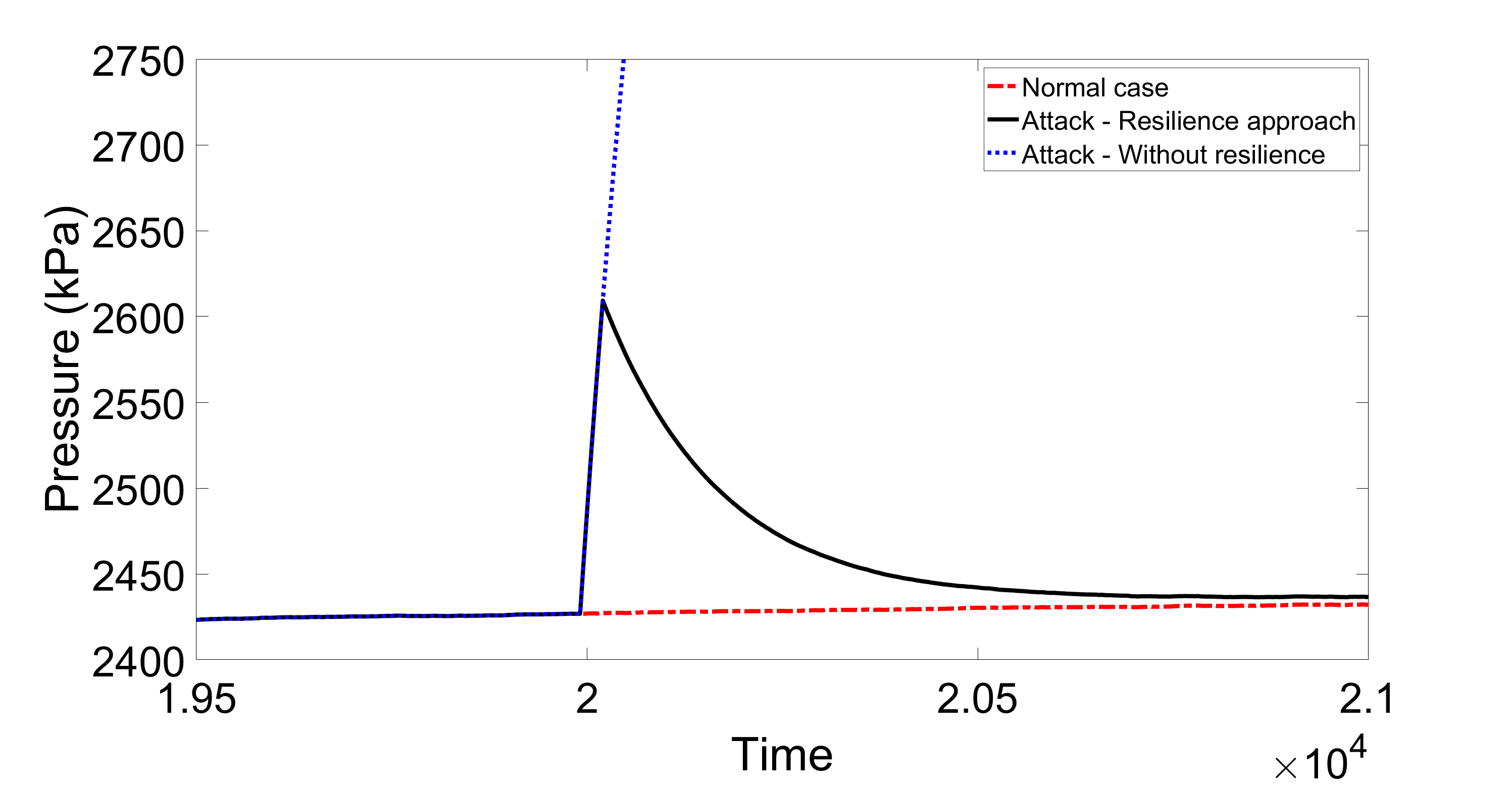}
  }
 \subfigure[\label{fig:max_res}]{
  \includegraphics[width=0.45\linewidth]{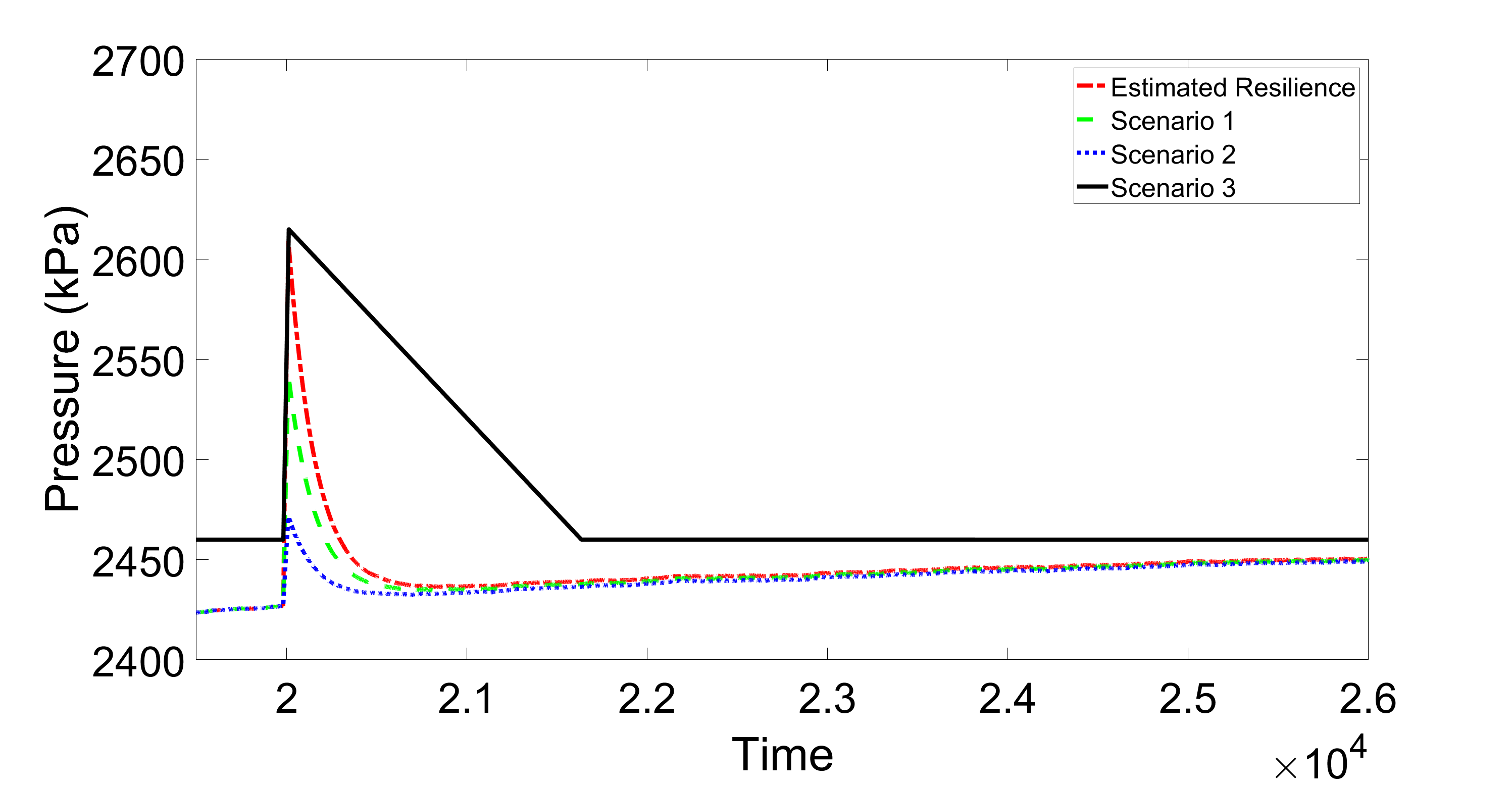}
 }
  \subfigure[\label{fig:min_no_res}]{
  \includegraphics[width=0.45\linewidth]{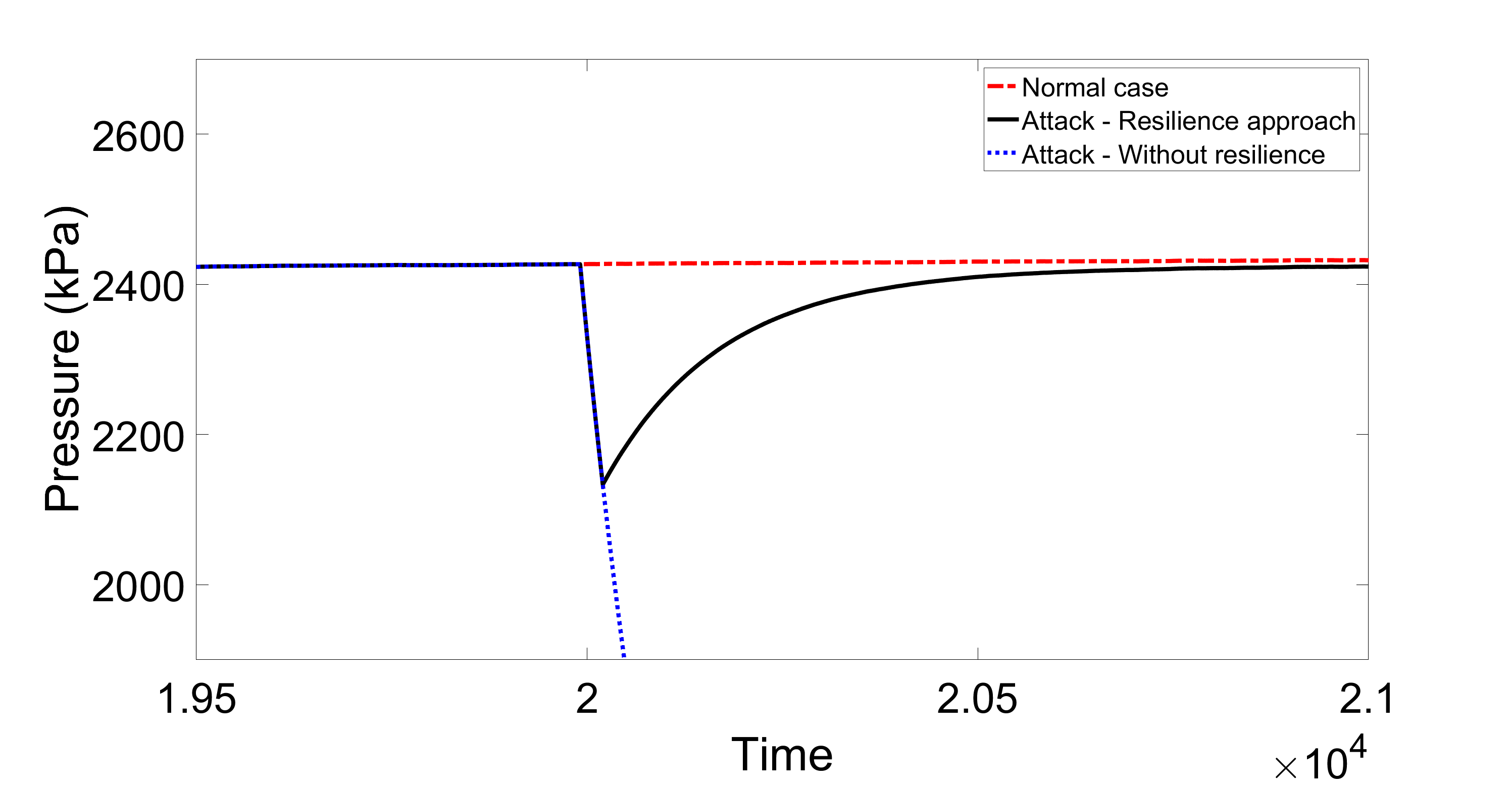}
  }
 \subfigure[\label{fig:min_res}]{
  \includegraphics[width=0.45\linewidth]{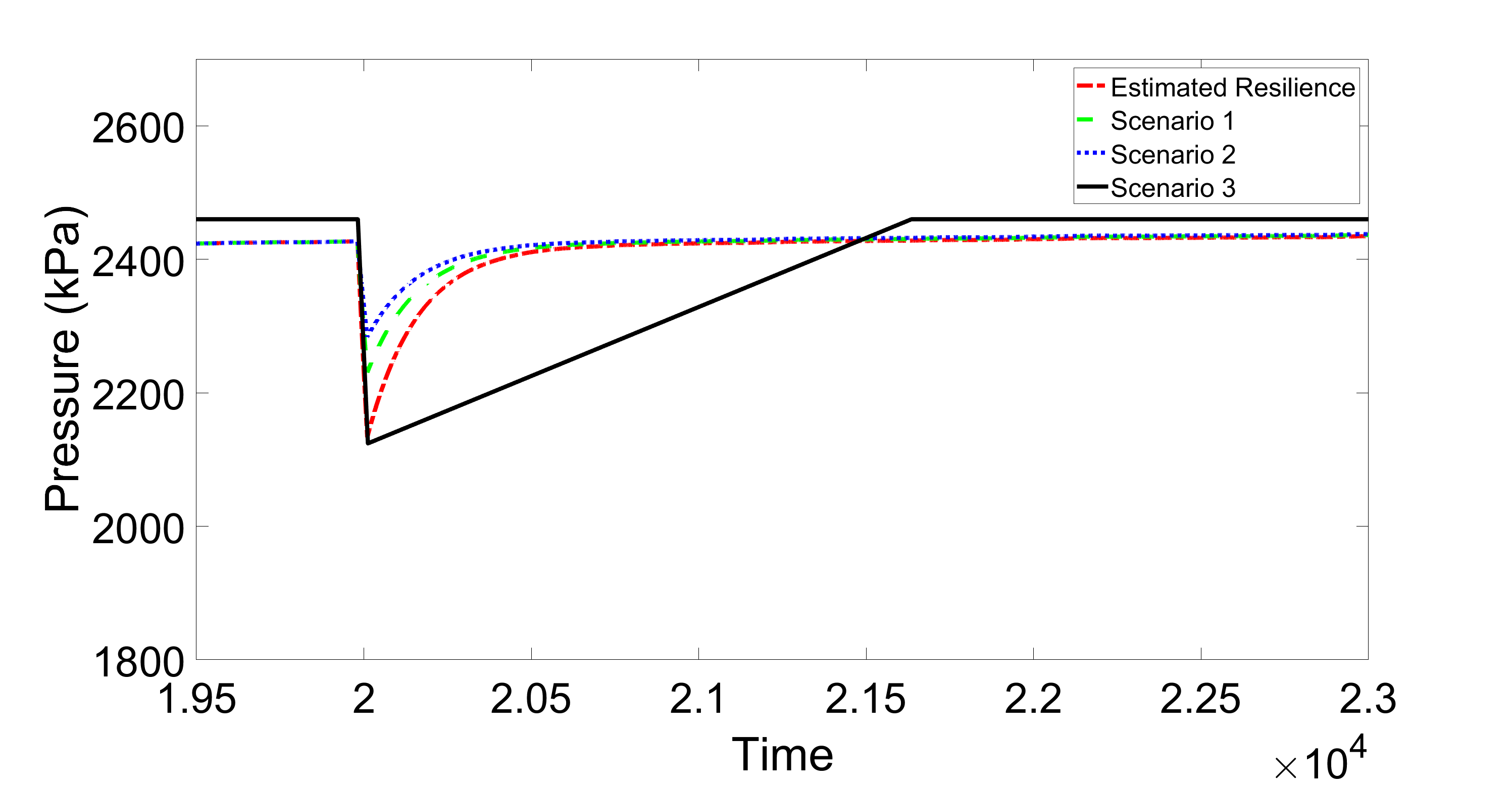}
 }
 \caption{(a) Resilient response vs normal case and attack without
   resilience for an adversary exploiting the maximum pressure
   threshold, (b) Resilience estimation using the proposed metrics vs.
   Monte Carlo simulation for adversaries in Table
   \ref{tab:scenarios_P_max}, (c) Resilient response vs normal case
   and attack without resilience for an adversary exploiting the
   minimum pressure threshold, (d) Resilience estimation using the
   proposed metrics vs. Monte Carlo simulation for adversaries in Table
   \ref{tab:scenarios_P_min}.
 }
\end{figure*}

\begin{table}[b]
\begin{center}

\begin{tabular}{| c | c | c | c |}
\hline
\textbf{Valve} & \multicolumn{3}{c|}{\textbf{Saturation}}  \\ \hline
\textbf{Scenario} & \textbf{\# 1} & \textbf{\# 2} & \textbf{\# 3} \\ \hline

$u_1$ & 100\% & 85\% & 70\%  \\ \hline
$u_3$ & 0\%   & 5\%  &  12.5\% \\\hline
\end{tabular}
\medskip
\caption{Malicious saturation level scenarios to exceed the system
maximum pressure.\label{tab:scenarios_P_max}}
\end{center}
\end{table}

\begin{table}[b]
\begin{center}

\begin{tabular}{| c | c | c | c |}
\hline
\textbf{Valve} & \multicolumn{3}{c|}{\textbf{Saturation}}  \\ \hline
\textbf{Scenario} & \textbf{\# 1} & \textbf{\# 2} & \textbf{\# 3} \\ \hline

$u_1$ & 0\%    & 20\%   &  30\%  \\ \hline
$u_3$ & 100\%  & 75\%  &  50\% \\\hline
\end{tabular}
\medskip
\caption{Malicious saturation level scenarios to exceed the system
minimum pressure.\label{tab:scenarios_P_min}}
\end{center}
\end{table}

\begin{table}[!b]
\begin{center}
\begin{tabular}{| l |  c | c | c | c | c | c| c| c| c |}
\hline
&  \multicolumn{5}{c|}{\textbf{Performance \& Stability}} \\ \hline\hline
\textbf{Scenario} & \textbf{KA} & \textbf{KR}  & \textbf{MD}  & \textbf{RL}  & \textbf{PR} \\ \hline\hline
\# 1 & 30 & 126 & 182  & 13640 &  99.93\% \\ \hline

\# 2 & 30 & 53 & 115 & 4517 & 99.98\%  \\\hline

\# 3 & 30 & 0 & 44 & 442 & 100\%  \\\hline\hline

Resilience Estimation & 30 & 1623 & 188 & 128240 & 99.80\%   \\\hline\hline
\end{tabular}
\medskip
\caption{Resilience evaluation for Tennesse Eastman problem. Scenarios described in Table \ref{tab:scenarios_P_max}.\label{tab:results_max}}
\end{center}
\end{table}

\begin{table}[!b]
\begin{center}
\begin{tabular}{| l |  c | c | c | c | c | c| c| c| c |}
\hline
&  \multicolumn{5}{c|}{\textbf{Performance \& Stability}} \\ \hline\hline
\textbf{Scenario} & \textbf{KA} & \textbf{KR}  & \textbf{MD}  & \textbf{RL}  & \textbf{PR} \\ \hline\hline

\# 1 & 30 & 491 & 293 & 66327 & 99.68\% \\\hline

\# 2 & 30 & 405 & 196 & 45900 &  99.78\% \\\hline

\# 3 & 30 & 339 & 143 & 34050 & 99.83\%  \\ \hline\hline

Resilience Estimation & 30 & 1623 & 302 & 277910 & 98.72\%   \\\hline\hline
\end{tabular}
\medskip
\caption{Resilience evaluation for Tennesse Eastman problem. Scenarios described in Table \ref{tab:scenarios_P_min}.\label{tab:results_min}}
\end{center}
\end{table}

The thresholds are essential to evaluate whether a system will be resilient or not. For each monitored variable with threshold restrictions, we should evaluate if the system will meet them or not considering the worst case adversary scenario. In this experimental work, we present the evaluation considering only the system pressure as the monitored variable. However, the process should be also repeated for the other variables.

\medskip

\noindent \textbf{Performance and Stability Metrics:} For the evaluation, we consider the resilience approach explained in Appendix \ref{append} and we want to measure how much this approach improves the resilience by calculating the defined metrics. To be resilient the system has to remain within the threshold for any adversarial input. In the case of the system pressure, we have a minimum and maximum threshold.

In this experimental work, we will consider two adversaries that want to exploit the pressure threshold. The first adversary makes the system exceed the maximum value and the second one the minimum. The configuration parameters for these adversaries are detailed in Tables \ref{tab:scenarios_P_max} and \ref{tab:scenarios_P_min} respectively. The scenarios use different saturation levels to represent adversaries' aggressiveness level in the adversary model.

As showed in Equation (cf. \ref{transferFunc}), the pressure can be obtained as $P= g_{21}.u_1 + g_{23}.u_3$. Hence, it depends on command inputs $u_1$ and $u_3$. In addition, $g_{21}$ has a positive sign, so, if we increase $u_1$, we will increase the pressure. On the contrary, $g_{23}$ has a negative sign, so we need to decrease $u_3$ value to increase the pressure.

Figures \ref{fig:max_no_res} and \ref{fig:min_no_res} compares the behavior of the system with the resilience approach facing both adversaries. In addition, it compares this response with the normal case behavior without attack and the attack case without resilience.

\medskip

To evaluate the resilience, we used the metrics defined in Section \ref{subsec:stab_perf} and we compare the behaviors against the defined threshold. The results for the maximum and minimum pressure threshold are showed in Tables \ref{tab:results_max} and \ref{tab:results_min} respectively.

The estimated resilience is obtained with the proposed metrics and it shows how the system will react during the absorb and recovery phase considering the worst case adversary. We can observe also that all the adversaries scenarios are included withing the resilience estimation and more aggressive adversaries, such as scenario $\#1$, produce a bigger decrease in resilience than a less aggressive such as scenario $\#3$.
We can observe this in Figures \ref{fig:max_res} and \ref{fig:min_res} that compare the estimated resilience with experimental Monte Carlo simulations for the scenarios in Tables \ref{tab:scenarios_P_max} and \ref{tab:scenarios_P_min} respectively.

\medskip

\noindent \textbf{Design and Structure Resilience:} Thereinafter, we discuss how to incrementally design a resilient TE system.

Design 1: The most basic design is a system with no automated controller feeding inputs to actuators. It is controlled, for example, manually by an operator or the actuators operate in a fixed way.

Design 2: Another option to create a basic design is a system that has no sensors and it works at open loop since the controller is not getting feedback from the physical process.

Designs 1 and 2 are not resilient to attacks. They are even not capable of correcting system failures because there is no controllability and no observability. For this reason, the metrics $R_A$, $R_S$, $R_C$ and $R_N$ are all zero.

Design 3: The previous design can be improved by providing basic observability and controllability with an automated controller capable of correcting errors in the process, non-redundant actuators, sensors and network.

This design is better that the previous ones because it ensures fault correction, i.e, it is capable of correcting non-malicious errors in the physical process. However, this design is still not resilient to attacks and the metrics $R_A$, $R_S$, $R_C$ and $R_N$ are all zero.

To improve the resilience, it is required to contemplate mechanisms to face the compromise of sensors, actuators, controllers or network links. If a system has more capabilities to restore the critical components than other system, then its more resilient. In the sequel, we provide examples to do this. These metrics do not represent the system security. Instead, a better resilience measure indicates that the system will react in a stable manner, recover with less effort and with less damage after an attack.

Design 4: We can improve Design 3 by adding a resilience approach such as the one in Appendix \ref{append}. This way, $R_N$ is increased in one.

Design 5 and 6: Adding diversified sensors and actuators it is possible to improve $R_S$ and $R_A$ respectively. For example, in a system with 4 actuators, it will be 2-actuator resilient if the system after removing any combination of 2 actuators has still the ability to find a control input that can take the system to an equilibrium state. This means that if the set $\Gamma$ which contains any combination of two not compromised actuators, i.e. $\{(a1,a2)$, $(a1,a3)$, $(a1,a4)$, $(a2,a3)$, $(a2,a4)$, $(a3,a4)\}$ will be 2-$R_A$ if all the systems defined for this set $\Gamma$ are controllable, i.e., $(A, B^{(a1,a2)})$, $(A, B^{(a1,a3)})$,  $(A, B^{(a1,a4)})$,  $(A, B^{(a2,a3)})$, $(A, B^{(a2,a3)})$ and $(A, B^{(a3,a4)})$ are all controllable.

Design 7: We can improve $R_C$ changing the valves for smart valves with an embedded controller integrated in the device. This option increases the resilience by adding redundant control outside the attacker domain, for example as in \cite{Rubio17ETT}. This will increase metric $R_C$ in one unit.

\section{Conclusion}
\label{sec:conclusion}

We have presented metrics to evaluate the resilience of a proposed approach. The proposed metrics are based on control theory performance and stability concepts; and on the design and structure of the system. A system with a better resilience indicates that the system will react in a stable manner, recover with less effort and with less damage after an attack. We evaluated the proposed metric using the Tennesse Eastman problem as a case study and we demonstrate the capabilities of the metrics to provide an upper bound for the worst case damage that an adversary may cause. The metrics also provide a mechanism to compare the resilience achieved by a particular approach or a whole system design, giving tools to evaluate during the system conception the best techniques to create a resilient design.

\bibliographystyle{plain}
\bibliography{main,main2,jga}

\appendix
\section{Appendix}
\label{append}

The resilience approach takes as an input a \gls*{cps} modeled by a transfer function and builds a resilient equivalent system capable of controlling the same physical process using a Switched Linear Control System. A switched system consists of a finite number of subsystems and a logical rule that orchestrates the switching between the subsystems. It may be modeled as follows:

\begin{equation}
  \label{eq}
  x_{k+1}=f_{\sigma(k)}(x_k, u_k)
\end{equation}

where $k \in \mathbb{Z}^+$ is the time interval, $x \in \mathbb{R}^n$ is the state, $u \in \mathbb{R}^p$ is the control input and $\sigma$ is the logical rule that orchestrates the switching between the subsystems. It means that $\sigma$ is a function $\sigma: \mathbb{Z}^+ \rightarrow \mathcal{I}$, where $\mathcal{I}= \{1,...,N\}$ contains the indexes of the subsystems. A subsystem is determined by a pair $(\mathcal{M}_i,\ \mathcal{G}_i)$ where $\mathcal{M}_i = \{A_i, B_i, C_i : i \in \mathcal{I}\}$ is the set of physical system models  and $\mathcal{G}_i = \{ V_i, E_i : i \in \mathcal{I} \}$ is the set of graphs that represent the network connections in the \gls*{cps}. Hence, $\sigma$ define a piece-wise switching signal that is a time-varying definition of the process model and the network graph that is activated at time $k$. The physical model activated at time $k$ is then defined by Equation~\eqref{eq2} as follows:
\begin{equation}
  \label{eq2}
    \left.
        \begin{array}{ll}
          x_{k+1}=A_{\sigma(k)}x_{k}+B_{\sigma(k)}u_{k} \\
          y_{k}=C_{\sigma(k)}x_{k}
        \end{array}
    \right.
\end{equation}
whose system communicates through a network determined by the connectivity
graph $\mathcal{G}_{\sigma(k)} = [V_{\sigma(k)}, E_{\sigma(k)}]$.
The approach aims at protecting the system from network adversaries working at the node level by modifying the controller model and at the network layers modifying the endpoint information. In the sequel, we describe the procedure to build the resilient system.

\medskip


\noindent \textbf{Step 1 (Models Design):} The first step to build the system is to design the physical models, i.e., create the subset of matrices $\mathcal{M}_i = \{A_i, B_i, C_i : i \in \mathcal{I}\}$  that will be activated at each time period.
The approach we propose is to design distributed controllers that modify in time the physical model they execute. The overall process is controlled by several independent controllers and altogether represent a decentralized controller, i.e., if at time $k$ it is activated the control model with matrices $A_i, B_i, C_i$ then there will be $j$ controllers with $j \in 1...o$ and each controller will use a set of matrices $A_{ij}, B_{ij}, C_{ij}$ where $ A_i = \bigcup\limits_{j=1}^{o} A_{ij}$, $B_i = \bigcup\limits_{j=1}^{o} B_{ij}$ and
$C_i = \bigcup\limits_{j=1}^{o} C_{ij}$. Hence, the controllers have available only parts of the overall information.

In the sequel, we describe how to derive the equivalent models starting from the initial transfer $G(s)$. The objective is to obtain different models expressed in a state-space model with the $A_{ij}, B_{ij}, C_{ij}$ matrices which can be combined to represent the system dynamics as in Equations~\ref{eq:ch3_state} and \ref{eq:ch3_output} and it allows deriving different sets of controllers capable of controlling the physical process.

\medskip

\begin{figure}[!t]
    \centering
    \subfigure[]{
        \includegraphics[width=0.5\textwidth]{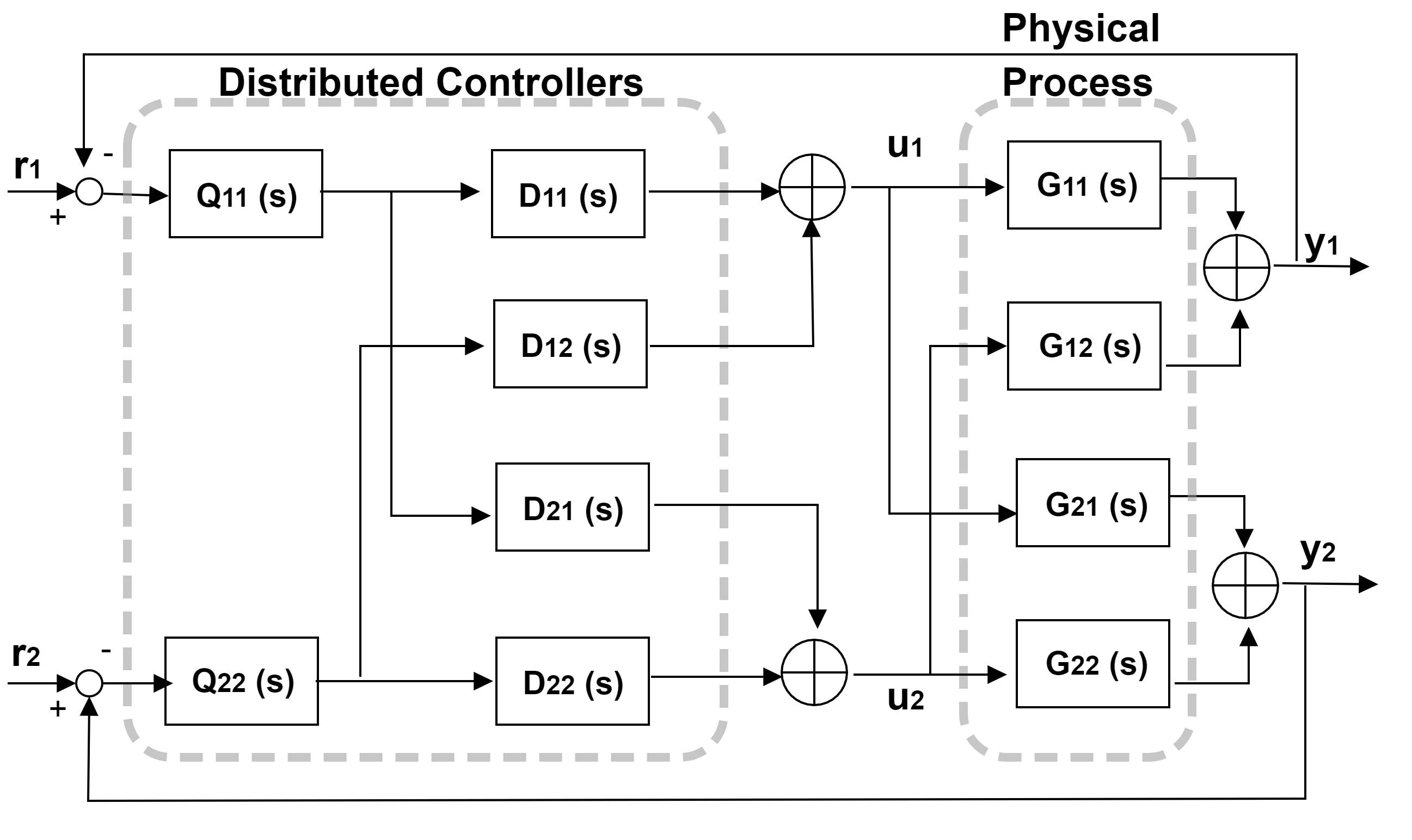}
     }
    \vspace{0cm}
    \subfigure[]{
        \includegraphics[width=0.25\textwidth]{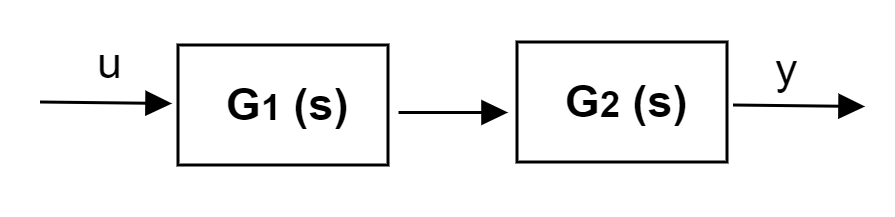}
     }
    \subfigure[]{
        \includegraphics[width=0.2\textwidth]{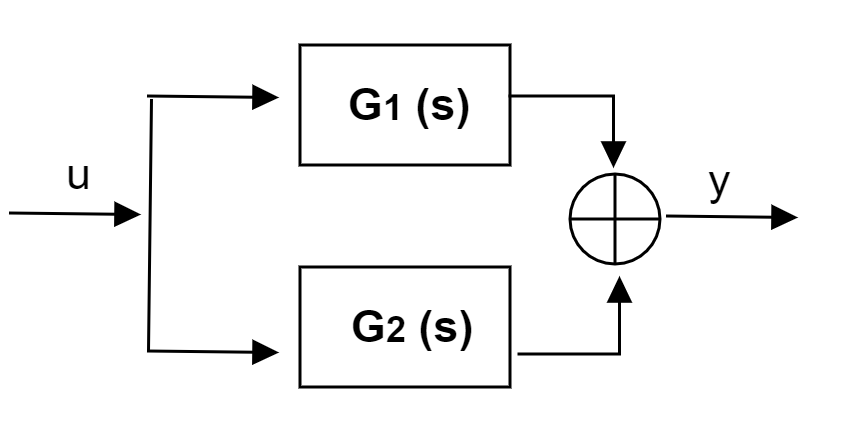}
    }
    \caption{Decentralized models (a) via serial decomposition (b) or parallel decomposition (c).}
    \label{fig:design}
\end{figure}

\noindent \textbf{Step 1.1:} To obtain the equivalent representation we will factorize the matrices applying techniques similar to the ones used by the different approaches for decentralized control design~\cite{liu_review_2019} \cite{wang_decoupling_2002}. It consists in combining a diagonal controller $Q(s)$ with a block compensator $D(s)$ in such a way that the controller perceives the process dynamics $G(s)$ as a set of independent processes as showed
in Equation~\eqref{dcp}:
\begin{equation}
  \label{dcp}
    G(s) \cdot D(s) = Q(s)
\end{equation}
where $D(s)$ and $Q(s)$ are both $n \times n$ matrices of transfer functions and $Q(s)$ is diagonal. Hence, the structure of the distributed controllers will be formed for $n$ controllers executing the $Q_{ii}$ transfer functions and each of these controllers is connected with $n$ controllers executing the $D_{ij}$ transfer function. In Figure~\ref{fig:design}(a), we show the structure for a $2 \times 2$ example.

To create this distributed design, the first step is to calculate $adjG(s)$ the adjudged matrix of G which is the transposition of the co-factor matrix of G.\\

\noindent \textbf{Step 1.2:} We build matrix $D(s)$ as follows.
For each column $\hat{J} = \{1,..,N\}$, we select a row $\hat{I}$ to set that element $d_{\hat{I}\hat{J}}$ in the matrix $D(s)$ to unity. It  is necessary to choose one for each column but not necessarily the diagonal ones.

After choosing the elements $(\hat{I},\hat{J})$ to be set to one, the matrix $D(s)$ can be completed as follows:
$$
d_{i\hat{J}} = \dfrac
        {adjG_{i\hat{J}}}
        {adjG_{\hat{I}\hat{J}}}
$$
where $adjG_{ij}$ is the $(i,j)$ element of $adjG(s)$ the adjugate matrix of G.

This process is repeated by selecting different rows $\hat{I}$ in order to obtain different distributed models. Hence, for a $n \times n$ process, there are $n^n$ possible choices of $D(s)$ since it depends on the possible positions to place the $1s$ values when building matrix $D(s)$. However, some of those choices can result in non-realizable systems. For example, if the adjudged matrix has a zero value in that entry. Thus, the configuration can be selected depending on the realizability.

\medskip

\noindent \textbf{Step 1.3:} $Q(s)$ is a diagonal matrix built using Equation~\eqref{dcp} and multiplying $G(s) \cdot D(s)$. Each matrix $D(s)$ gives, as a result, a different matrix $Q(s)$.

\medskip

\noindent \textbf{Step 1.4:}
Due to realizability restrictions, it is possible to have matrices $D$ with many elements equal to $0$, which reduces the number of possible generated models. In this case, it is possible to generate other equivalent models using transfer function decomposition techniques.

\medskip

\noindent \textbf{Step 1.4.1 (Serial Decomposition):}
A transfer function $G(s)$ may be decomposed in transfer functions that multiply together as showed in Figure~\ref{fig:design} (b). Hence, $G(s) = G_1(s).G_2(s)$. This decomposition is commutative and it is possible to  generate combinations of the different factors to create the distributed transfer functions.
This can be applied at the level of transfer functions as well as factoring the original transfer function in its poles and zeros representation as follows:
\vspace{-0.2cm}
\begin{equation}
  \label{serial}
    G(s) = k \prod_{i=1}^{N} \frac{s-z_i}{s-p_i}
\end{equation}
where the denominator coefficients $p_i$ are the poles, the
numerator $z_i$ are the zeros of the transfer function
and  $k$ is the gain term. This mechanism allows generating different partitions of matrices $Q(s)$ and $D(s)$.

\medskip

\noindent \textbf{Step 1.4.2 (Parallel Decomposition):} In this case, the transfer function $G(s)$ is decomposed into a sum of terms as showed in Figure~\ref{fig:design} (c). Hence, $G(s) = G_1(s)+G_2(s)$. This can be done with a technique called partial fraction decomposition that finds the residues and poles. The terms are as follows:
\vspace{-0.2cm}
\begin{equation}
  \label{parallel}
    G(s) = k + \sum_{i=1}^{N} \frac{r_i}{s-p_i}
\end{equation}
where the denominator coefficients $p_i$ are called the poles of the transfer function, the numerator $r_i$ is the residue of pole $ p_i$ and $k$ is a constant.
Hence, after applying this technique to a $d_{ij}$ transfer function, we will obtain a family of $d^{t}_{ij}$ functions that can be added to obtain the original $d_{ij}$ function.

This mechanism allows generating a different distribution of compensators matrices $D(s)$.

\medskip

\medskip

\noindent \textbf{Step 1.5:} After calculating the sets of matrices $D(s)$ and $Q(s)$, it is possible to take each $d_{ij}$ and $q_{ij}$ entry to calculate its corresponding matrices A, B and C using the procedure to transform a transfer function into a state space model.

\medskip


\noindent \textbf{Step 2 (Network Design):} In this section, we analyze how to design the network connectivity graph $\mathcal{G} = [V, E]$ for each of the physical models created in Step 1.

\medskip

\noindent \textbf{Step 2.1:} The transfer functions in $Q(s)$ are controllers that take one input and send one output. Each of them will be executed in one node. For notation, if a node $v_q$ executes the controller $q_{ii}$ then we will call it $v_{q_{ii}}$.

The $d_{ij}$ and $d^{t}_{ij}$ elements take the output of the $q_{jj}$ element to make their calculations and produce an output control signal. Each $d_{ij}$ will be executed in one node $v_d$ and the notation will be  $v_{d_{ij}}$ to express that the node $v_d$ executes the transfer function $d_{ij}$.
s
The network contains also a set of sensor nodes $v_s$ and a set of actuator nodes $v_a$. If the sensor measures the variables of $G_{ij}$, then the notation will be $v_{s_{i}}$. In a similar way, $v_{a_{j}}$ represents the actuator that applies the control input $j$.

Hence, the set of nodes $V$ in the graph $\mathcal{G}$ contains the nodes $v_q$, $v_d$, $v_s$ and $v_a$. In the system, there are also network devices, such as routers and switches. However, we are not explicitly including them in the design as we assume a traditional use of them.

\medskip

\noindent \textbf{Step 2.2:} The set of edges $E$ will be defined from the matrices $D(s)$ and $Q(s)$ according to the following four main rules: (1) $(v_{q_{jj}},v_{d_{ij}}) \in E$; (2) $(v_{d_{ij}}, v_{a_{i}})\in E$; (3) $(v_{d^{t}_{ij}}, v_{a_{i}})\in E$; (4) $(v_{s_{i}}, v_{q_{ii}})\in E$.
An example can be observed in Figure \ref{fig:design}(a) where according to the rule (1) the component $q_{11}$ is connected to $d_{11}$ and $d_{21}$. In addition, the output of $q_{22}$ should be sent to $d_{12}$ and $d_{22}$. Due to rule (2), the output of components $d_{11}$ and $d_{21}$ are combined to create the command $u_1$ that should be received by actuator $a_1$. In a similar manner, it is created the command for actuator $a_2$. Rule (3) is the equivalent to rule (2) when parallel decomposition is applied. In this particular case, it does not apply. Finally, rule (4) indicates that the sensor $s_1$ and $s_2$ measure the data that should be sent to components $q_{11}$ and $q_{22}$ respectively.

\medskip

\noindent \textbf{Step 2.3:} To coordinate the system, there will be an orchestrator, physically located in the SDN controller. The orchestrator has the following responsibilities.

\begin{enumerate}

\item \textbf{Choose a key for the model selection}. There are $\mathcal{I}= \{1,...,N\}$ possible subsystems to activate and the orchestrator chooses in a random manner a key $K_1$ which will be used to select the next model to activate using a hash function as follows $hash(K_1,j)\ mod\ N$ where $j$ is the switching interval.
The common sharing of $K_1$, $j$ and $N$ allows each device to compute the next active model in a distributed manner. The key is renewed periodically using one of the existing approaches for key generation and distribution such as \cite{8403548}.

\item \textbf{Coordinate the network configuration transformation.} Each component will change its network configuration in each switching period of the physical model. To do this, each device gets  a real IP address (RIPA) and a virtual IP address (VIPA). The RIPA is used for management purposes making the network configuration transformation transparent to administrators. The VIPA is used to communicate the data packets of the \gls*{cps}, i.e., the hosts communicate with another host using their VIPAs. In addition, VIPAs change periodically and synchronously in a distributed fashion over time. In every transformation interval, the hosts will be associated with a unique VIPA.

 The VIPA  transformation is managed by the SDN devices by selecting an address from the unused address space.
Each host will be allocated an IP address ranges to choose the VIPAs and they are selected using a hash function from the designated ranges. Since the VIPAs are chosen from the assigned network sub-nets, there is no need to do a routing update advertisement for internal routers. In addition, SDN devices will forward packets from old connections until the session is terminated or expired.

Each SDN device is responsible for the management of the hosts in one or more sub-nets. The VIPAs selection is done in a similar way to the physical model selection. It uses a hash function and a secret random key to guarantee unpredictability. If there are $p$ available VIPAs for a host, then the SDN device can compute the index of the VIPA for the switching interval $j$ as $hash(K_2,j)\ mod\ p$. The SDN controller is responsible for the management of the SDN devices and the key $K_2$ distribution.

\item \textbf{Coordinate the transformation time}. The orchestrator has to choose and coordinate the switching in a master-slave mode. It requires a distributed timing synchronization that ensures the achievement and maintenance of a common time for all the nodes of the network. Many proposals have already work in solving this type of issues~\cite{1316761}.

\end{enumerate}

\medskip

\noindent \textbf{Step 3 (Switching Function Design):} Finally, it is required to design the switching function $\sigma$ which indicates when to change the activated subsystem. It can be demonstrated that from the physical point of view, it is possible to use an unrestricted switching signal, this means that there is no minimum switching time required since the proposed subsystem share a Common Quadratic Lyapunov function by design. Hence, this ensures the stability of the proposed switched linear system. However, in this type of system, the physical part is coupled with the cyber components and for this reason, the switching must be done considering the correct behavior of the cyber layer, for example, a switching time that allows the network devices to update correctly the routing tables.

\end{document}